# Detecting Pulmonary Coccidioidomycosis (Valley fever) with Deep Convolutional Neural Networks

Jordan Ott[1], David Bruyette[2], Cody Arbuckle[2], Dylan Balsz[2], Silke Hecht[3], Lisa Shubitz[4], Pierre Baldi[1, 5, *]
jott1@uci.edu, david@anivive.com, cody@anivive.com, dylan@anivive.com, shecht@utk.edu,
lfshubit@email.arizona.edu, pfbaldi@ics.uci.edu
[1] Department of Computer Science, University of California, Irvine, USA
[2] Anivive Lifesciences, Inc. , Long Beach, California, USA
[3] College of Veterinary Medicine, University of Tennessee, Knoxville, TN, USA
[4] Valley Fever Center for Excellence, The University of Arizona, Tucson AZ, USA
[5] Institute for Genomics and Bioinformatics, University of California, Irvine, USA

## Abstract

Coccidioidomycosis is the most common systemic mycosis in dogs in the southwestern United States. With warming climates, affected areas and number of cases are expected to increase in the coming years, escalating also the chances of transmission to humans. As a result, developing methods for automating the detection of the disease is important, as this will help doctors and veterinarians more easily identify and diagnose positive cases. We apply machine learning models to provide accurate and interpretable predictions of Coccidioidomycosis. We assemble a set of radiographic images and use it to train and test state-of-the-art convolutional neural networks to detect Coccidioidomycosis. These methods are relatively inexpensive to train and very fast at inference time. We demonstrate the successful application of this approach to detect the disease with an Area Under the Curve (AUC) above 0.99 using 10-fold cross validation. We also use the classification model to identify regions of interest and localize the disease in the radiographic images, as illustrated through visual heatmaps. This proof-of-concept study establishes the feasibility of very accurate and rapid automated detection of Valley Fever in radiographic images.

* Corresponding author: pfbaldi@uci.edu

# 1. Introduction

Coccidioidomycosis, or "Valley Fever", is a disease found in dogs with the ability to infect humans. Currently found in the southwestern United States, the disease is expected to spread with warming climates in the coming years. The disease can be detected through radiological examination of the lungs. In order to aid in the identification and diagnosis of the disease we leverage deep learning classification models. These models have been applied to medical imaging especially in human clinical settings. In this paper we apply deep learning models to veterinary medicine and radiology images of healthy and infected dogs.

**1.1 Background on Valley Fever**

The disease results from infection by the dimorphic, saprophytic fungal organism Coccidioides immitis or Coccidioides posadasii. Infection by Coccidioides spp. occurs in a wide range of host species, including humans, a myriad of domestic and exotic mammals, and, rarely, reptiles (Shubitz, 2007). Coccidioidomycosis is endemic in semiarid regions of California, Arizona, New Mexico, Texas, and Northern Mexico and semiarid regions in South America. Across this region, temperature and precipitation influence the extent of the endemic region and number of Valley Fever cases. Climate projections for the western US indicate temperatures will increase and precipitation patterns will shift, which may alter disease dynamics. Using a climate niche model derived from contemporary climate and disease incidence data with projections of climate from Earth system models, researchers have assessed how endemic areas will change during the 21st century. By 2100 in a high warming scenario, the model predicts the area of climate-limited endemicity will more than double, the number of affected states will increase from 12 to 17, and the number of Valley fever cases will increase by 50%. The Valley Fever endemic region will expand north into dry western states, including Idaho, Wyoming, Montana, Nebraska, South Dakota, and North Dakota. Precipitation will limit the disease from spreading into states farther east and along the central and northern Pacific coast (Gorris, 2019).

**1.2 Incidence of Valley Fever**

Inhalation is the most common route of infection in both animals and humans and typically occurs after fungal hyphae have desiccated and matured into arthroconidia that are easily aerosolized. The arthrospores are inhaled and dispersed along the bronchial tree. Here, they undergo structural transformation into spherules, which enlarge and undergo endosporulation. Eventually, new endospores are released into the surrounding tissue and the cycle continues until the host is able to mount an appropriate immune response. Direct cutaneous inoculation of arthrospores has been reported, but these instances are rare and typically result in local granuloma formation (Tortorano, 2015).

Infections caused by Coccidioides may be subclinical or result in severe illness and death. Dissemination to extrapulmonary organs is a potential sequela in some animals. One study indicated that 80% of dogs develop primary pulmonary infection whereas 20% develop disseminated disease Davidson, 1994). Pulmonary coccidioidomycosis is characterized by clinical signs that include chronic cough, lethargy, and respiratory distress. Radiographic pulmonary lesions may be characterized by interstitial to nodular patterns, with hilar lymphadenopathy (Mehrkens, 2016).

Human infection has steadily increased over the last 10 years, and coccidioidomycosis is considered a reemerging infectious disease. Dogs come into close contact with soil, because of a closer proximity to the ground and digging behavior, and may provide a clearer delineation of broadly defined endemic regions. In this sense, rapid recognition of Coccidioides infection in dogs in a clinical setting may benefit human public health as well as minimize morbidity and owner expense.

**1.3 Deep Learning for Biomedical Imaging**

Deep learning (Baldi, 2021), a rebranding of neural networks, is becoming prevalent in the biomedical imaging domain (Baldi, 2018). Recent examples include colonoscopy screening (Urban, 2018), cardiovascular disease

(Wang, 2017), spinal metastasis (Wang, 2017), classifying genetic mutations (Chang, 2018), skin cancer detection (Esteva, 2017), and counting hair follicles (Urban, 2020) to name just a few. Additionally, previous studies have utilized machine learning and convolutional neural networks to evaluate thoracic radiographs in man (Tang, 2018). There has been considerably less work in applying deep learning methods to veterinary medicine in general, and thoracic radiographs in particular. These applications come with their own challenges, not the least of which is the potential for greater variability in anatomy and dimensions, for instance among dogs.

In this study, we apply deep convolutional neural networks in the evaluation of a single well-defined medical condition (pulmonary coccidioidomycosis) in dogs.

## 2. Materials and Methods

### 2.1 Radiographs of normal and abnormal dogs

The challenge study was performed at Colorado State University under an Institutional Animal Care and Use Committee (IACUC) approval. The normal dogs consisted of two groups. One were purpose bred laboratory dogs used as the control group in the development of a challenge model of canine coccidiomycosis. The dogs were equally distributed between males and females and were all of the same age (approximately twelve months) and weight. The second group consisted of normal, healthy pet dogs of various ages and weights with thoracic radiographs reviewed by a veterinary board-certified radiologist.

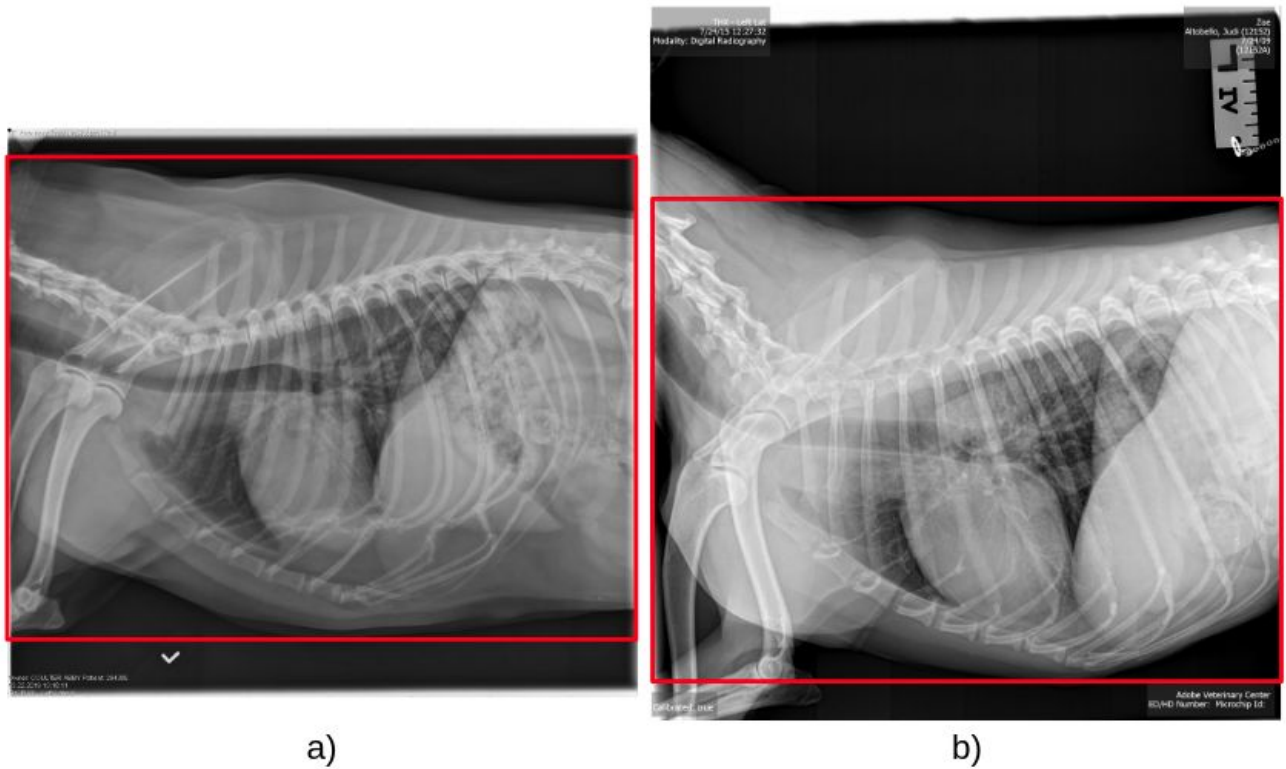

Figure 1. Images of canine subjects. Red bounding boxes represent cropped regions during preprocessing. a) Healthy subject; b) Subject with coccidioidomycosis.

The diseased dogs consisted of two groups. One were purpose bred laboratory dogs that were infected with varying doses of live spores of Coccidiodies posadasii as part of the aforementioned challenge study. The dogs were all of the same age (approximately twelve months) and weight. Infection was confirmed by histopathology and fungal lung and lymph node cultures. The second group consisted of pet dogs living in the endemic area with a confirmed diagnosis of coccidioidomycosis. All cases were evaluated at a single veterinary specialty hospital in Phoenix, Arizona.

The radiographic images were captured using commercially available veterinary digital imaging systems. Recorded images were deidentified and given a unique id number indicating normal or diseased, prior to being evaluated in the study. Right lateral (RL), left lateral (LL), and ventrodorsal (VD) views were obtained and the images annotated with the dogs unique ID number, date and view. The dataset consists of 106 infected dogs and 240 healthy dogs. The digital captures of these subjects produced 1,186 total images, 12 of which were marked as irrelevant and as a result discarded. The discarded images captured other body regions such as the front or hind legs, and critically, did not include the lungs or chest region, making the identification of the disease impossible in these images. After filtering irrelevant images, the remaining images were of 769 healthy subjects and 405 infected subjects, for a total of 1,174 images. All images were resized to 500 by 400 pixels for computational efficiency and to produce uniform image sizes across the dataset. A constant image size is required for the classification models employed in this study, which are described in the following section.

All images underwent a manual preprocessing phase, where each image was subject to cropping by a human expert. This process ensured the removal of many spurious features such as textual and label markings. Examples of selected crops are shown in Figure 1.

**2.2 Neural Networks**

Fully-connected, feedforward neural networks (NNs) receive a single vector representing the raw data at the input layer. This input is then transformed through a series of weighted connections to hidden layers that perform non-linear operations, before being routed to an output layer for the purposes of classification, regression, and other tasks. In these architectures, each node in a layer is connected to every node in the following layer. Though powerful, fully-connected feedforward networks can require a large number of parameters and are not ideally suited for processing images and recognize features in ways that are invariant under translations and other deformations. For computer vision applications, most of the time Convolutional Neural Networks (CNNs) (Fukushima, 1982; LeCun, 1998; Baldi, 1993) are a better choice. CNNs use weight sharing to reduce the number of parameters and convolve the image with arrays of local filters that are learnt from the data, providing a basis for building equivariance/invariance to translation and other geometric transformations in the overall response.

In this work five different CNN architectures are implemented, trained, and compared to classify the disease status of canine subjects: Inception (Szegedy, 2016), MobileNet (Sandler, 2018), ResNet (He, 2016), VGG (Simonyan, 2014), as well as a relatively shallow four-layer network. The first four of these architectures have been applied to a variety of image based problems in the literature.

# 3. Experiments

**3.1 Data Augmentation**

CNNs require large amounts of training data, specifically when the input space is large and the model has a high number of parameters. Insufficient amounts of data can lead the network to overfit, where it performs very well on the training set but fails to generalize properly to outside examples.

In order to combat overfitting, one basic approach is data augmentation where the original data is used to create additional training examples through various transformations. Data augmentation in this study was performed using rotations, translations, reflections, noise addition, and zooming. Each batch presented to the neural network during training was first normalized by color channel, by subtracting the mean value and dividing by the standard deviation. Then each normalized image was rotated by a random amount chosen uniformly in the 0 -- 15 degrees range, translated left or right, and up or down, by a random amount chosen uniformly in the 0--10% range of the original width and height, horizontally flipped, perturbed by adding Gaussian noise with mean 0 and standard deviation 0.1 to each pixel, and cropped by zooming in on the center of the image by a random uniform amount in the 0 -- 10% range. Starting from the small dataset of 1,174 images, we randomly augmented each batch of training examples, effectively increasing the number of available examples by roughly two orders of magnitude.

**3.2 Neural Network Training**

Of the architectures used in this study, Inception, MobileNet, ResNet, and VGG were selected based on their popularity and performance on benchmark datasets. The fifth model, the four layer shallow network, was selected to serve as a baseline architecture for comparison purposes. Inception, MobileNet, ResNet, and VGG have the option to use a set of weights that were previously trained on the ImageNet dataset (Deng, 2009). This process is known as transfer learning or pre-training, where weights trained on one task are then reused for a different task. When using the ImageNet weights, the primary convolutional layers can be frozen, and only new output layers are trained. The option, of using pre-trained weights was treated as a hyperparameter to be optimized, and thus we trained both networks that were initialized randomly and networks that were initialized with pre-trained weights.

All networks were trained for 100 epochs using an early stopping condition if the validation accuracy did not improve in the most recent 15 epochs. All models were trained using 10-fold cross-validation. This process partitions the data evenly into 10 distinct subsets and each fold trains a model using 9 subsets and holds one subset out for testing purposes. The partitioning was stratified to ensure a constant ratio of representation amongst healthy and infected examples - roughly 66% healthy and 34% diseased, in keeping with the overall image ratio. As the classes are unbalanced the network may learn to favor the majority class thus increasing its accuracy simply by chance. Class weighted penalties in the loss function can alleviate this issue. The use of such loss are explored in the hyperparmeter options in the next section. Ten-fold cross validation requires each model to be trained ten distinct times (re-initializing network parameters each time) and ensures different subsets of the data are used for training and testing. During a given fold, images in the training set never appear in the validation set and images in the validation set are not present in the training set. All models were implemented in Keras with a Tensorflow backend.

Table 1: Hyperparameter Space. The hyperparameters from the best performing architecture variants are shown in their respective columns.

| Name | Range | Parameter Type | Inception | MobileNet | ResNet | Shallow | VGG |
|---|---|---|---|---|---|---|---|
| Dropout | 0., 0.25 | Continuous | 0.28 | 0.29 | 0.31 | 0.44 | 0.21 |
| Learning Rate | 0.00001, 0.01 | Continuous | 0.0006 | 0.0096 | 0.0071 | 0.0025 | 0.0056 |
| Learning Rate Decay | 0.5, 1. | Continuous | 0.93 | 0.82 | 0.98 | 0.93 | 0.86 |
| Loss Penalty | Yes, No | Choice | No | No | No | No | Yes |
| Number of Layers | 0, 2 | Discrete | 0 | 0 | 1 | 1 | 0 |
| Number of Nodes | 128, 512 | Discrete | - | - | 152 | 180 | - |
| Optimizer | Adam, SGD, RMSProp | Choice | Adam | SGD | SGD | RMSProp | RMSProp |
| Weights | ImageNet, Random | Choice | Random | ImageNet | ImageNet | Random | ImageNet |
| Train base | Yes, No | Choice | Yes | Yes | Yes | No | No |

### 3.3 Hyperparameter Optimization

When implementing neural networks there are a number of choices - hyperparameters - that must be set prior to training. Choosing hyperparameters arbitrarily can lead to suboptimal results. To address this, we conducted neural network optimization via a random hyperparameter search using SHERPA (Hertel, 2020), a Python library for hyperparameter tuning. The random search algorithm has the advantage of making no assumptions about the structure of the hyperparameter search problem and is ideal for exploring a variety of settings.

We detail the hyperparameters of interest in Table 1, as well as the range of available options during the search. The hyperparameters of interest consisted of the amount of dropout, learning rate, learning rate decay, nodes per layer, the number of dense layers added at the end, a class weighted loss penalty, and the optimizer. As mentioned in the previous section, additional hyperparameters related to the use of pre-trained weights were also considered. The Inception, MobileNet, ResNet, and VGG networks had the option to use weights pretrained on the ImageNet dataset or a standard random initialization. Finally, the option of training the base convolutional layers was tested as well. The use of ImageNet weights and training the base layers are not mutually exclusive. For example, one model may start with the pretrained ImageNet weights, as an initialization, and training may proceed for both the top layers and the base layers.

Twenty-five different hyperparameter settings were tried for each network architecture, for a total of 125 models. Each trial was trained on a single split of the data, then following the hyperparameter search the best trials from each model were retrained in full using the 10 fold cross validation method. The hyperparameters from the best performing models are displayed in the corresponding columns of Table 1.

## 4. Results

The five architectures examined in this study were each run with twenty-five different combinations of hyperparameters. The best hyperparameter configuration from each architecture is shown in Table 1. Following the hyperparameter search, the best performing variant (determined according to validation accuracy) of each architecture was selected for further analysis. The validation accuracies during training are shown in Figure 2. All architectures demonstrated success reaching classification performance well above chance, represented by the dashed line. From the accuracy curves over time only MobileNet shows degradation at the later stages of training. The solid lines in Figure 2 show the mean validation accuracies while the shaded region represents one standard deviation of the 10 separate cross validation folds.

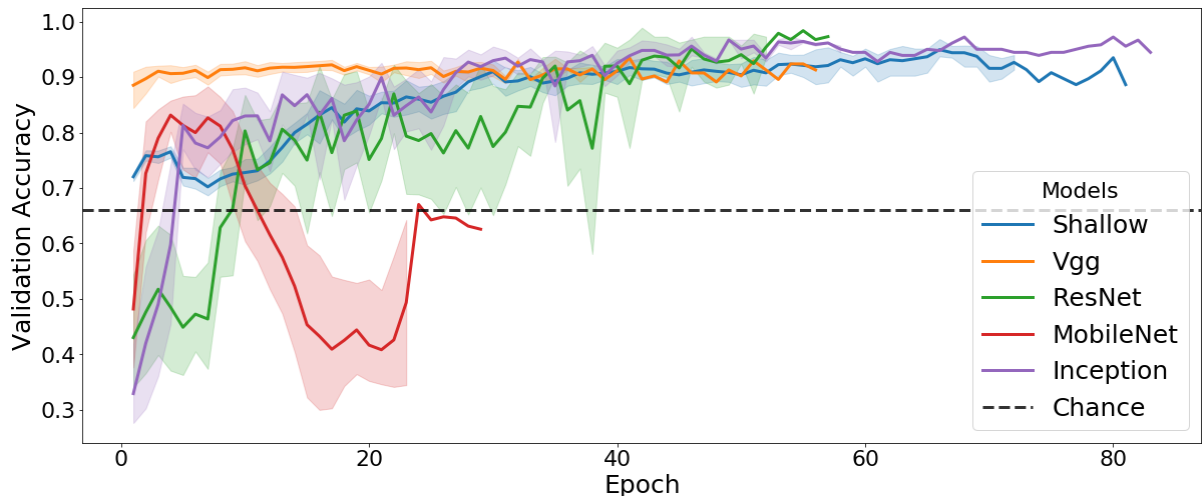

Figure 2: Validation accuracy during training. The legend indicates the corresponding model and the dashed line represents the accuracy from chance (around 66%).

Next we analyzed the relationship between the model's false positive and true positive rate in Figure 3. All models perform well above chance, with the worst performing model, MobileNet, achieving a mean AUC of 0.92. Examining the 10 fold cross-validation results shows the ResNet architecture as the dominant model. It achieves the highest mean AUC and lowest standard deviation over the 10 folds at 0.991 and 0.006 respectively.

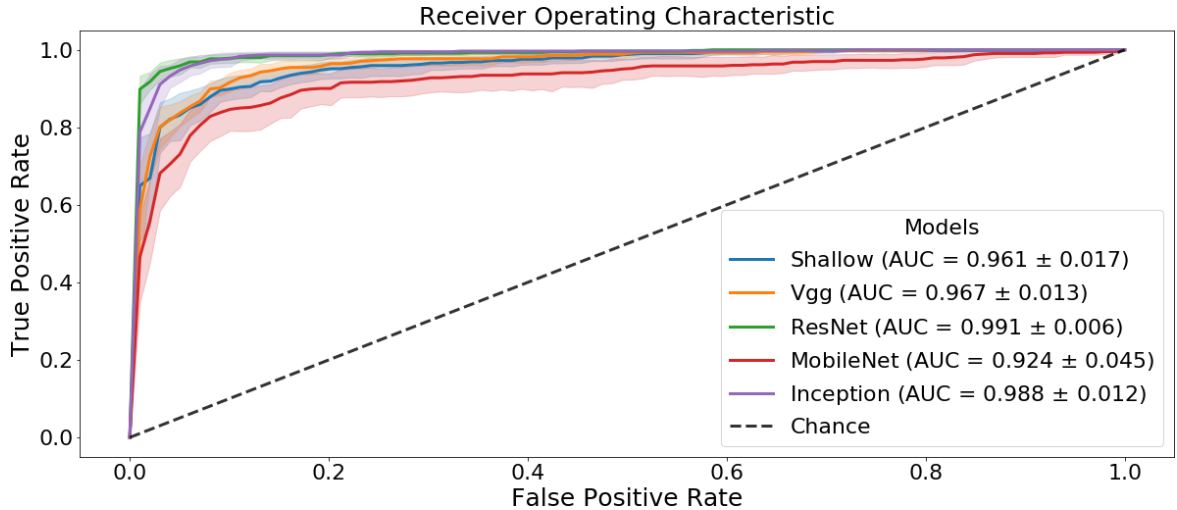

Figure 3: Receiver Operating Characteristic (ROC) curves displaying the relationship between the model's false positive rate and true positive rate. The legend indicates the corresponding model as well as the mean and standard deviation of the AUC (Area Under the Curve) obtained by 10-fold cross validation.

These results indicate the CNN models employed in this study can successfully solve this classification task. Having identified ResNet as the best performing model, we next analyzed its ability to locate important features related to coccidioidomycosis through the use of class activation maps. This exploration will ensure the model learns features relating to coccidioidomycosis and not spurious artifacts of the radiographic images.

Table 2: Result metrics showing sensitivity, specificity, positive predictive value, and negative predictive value.

|  | Sensitivity | Specificity | Positive Predictive Value | Negative Predictive Value |
| --- | --- | --- | --- | --- |
| Inception | 0.9056 | 0.9753 | 0.9485 | 0.9603 |
| VGG | **0.9254** | 0.8846 | 0.7836 | **0.9645** |
| ResNet | 0.8462 | **0.9964** | **0.9908** | 0.9364 |
| Shallow | 0.8021 | 0.9702 | 0.9216 | 0.9178 |
| MobileNet | 0.7688 | 0.8708 | 0.8001 | 0.9011 |

## 4.1 Visualizations

Neural networks are generally considered "black-box methods", which makes model interpretability an issue. However, certain methods like Class Activation Maps (CAM) project information back to the input space, allowing a visual understanding of the models decision (Zhou, 2016). CAM gives convolutional networks tremendous interpretability despite being trained on image-level labels. CAM requires the use of a global average pooling layer (Lin, 2013), which is added to the last convolutional layer of the models used in this study. The Keras Visualization Toolkit (Kotikalapudi, 2017) is used to produce CAM results. Using CAM, we are able to visualize what regions of input images the network attends to when making its classification prediction. This allows one to ensure the network is learning features directly related to coccidioidomycosis and not other circumstantial features contained in the images (e.g. borders or image annotations).

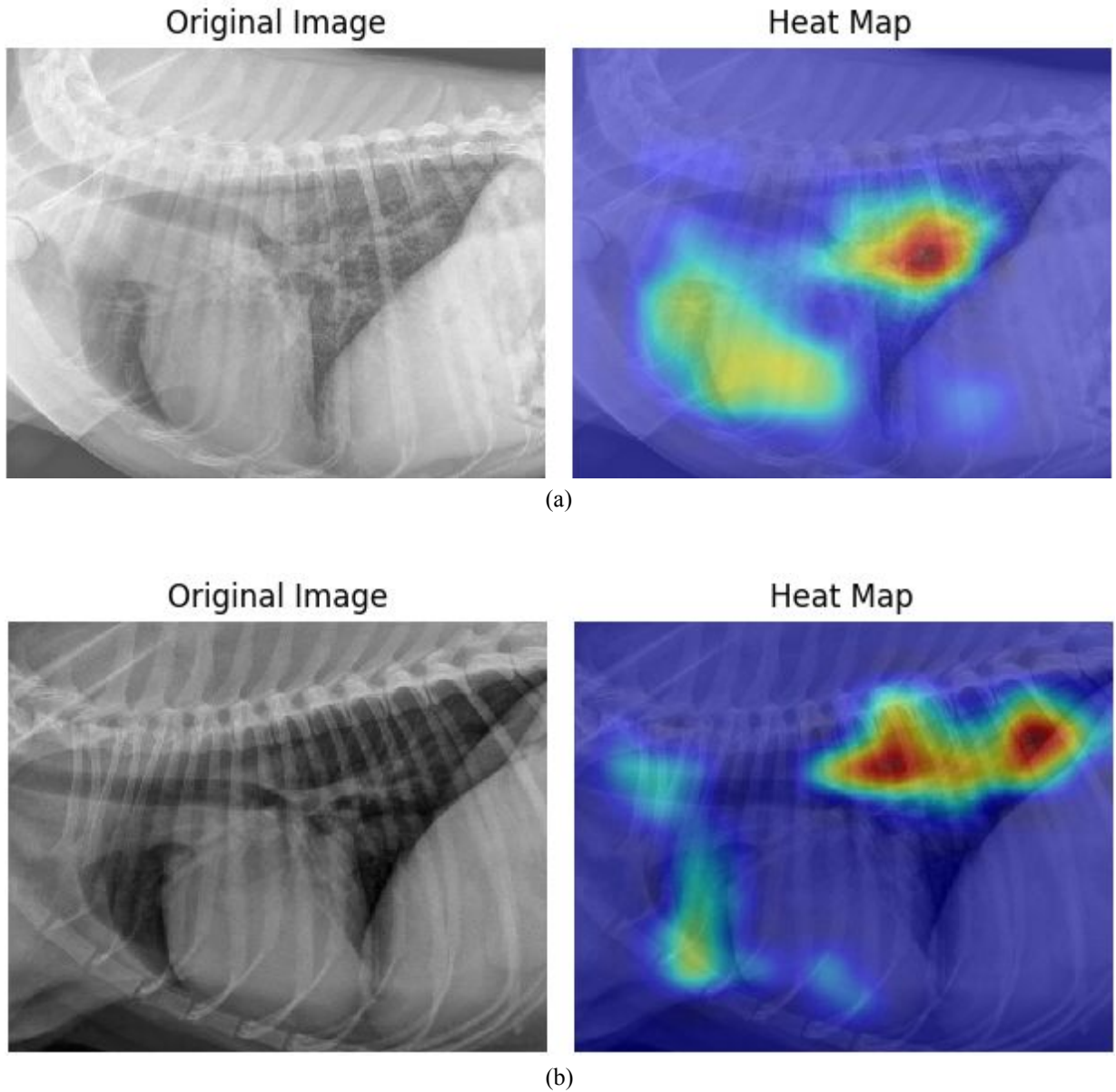

(a)

(b)

Figure 4: Original image fed to convolutional network and the class activation map produced by the ResNet's classification prediction. a) A canine subject with coccidioidomycosis correctly classified by the network. b) A healthy canine subject correctly classified by the network.

Shown in Figure 4a and 4b are correctly classified diseased and healthy subjects, respectively. The right panel shows the original images and the left panel displays the CAM results from the ResNet models classification. The CAM images are interpreted like heat maps, where the more red the region, the higher weight the network associates with those features when making its prediction. From the CAM images shown in Figure 4, one can verify the network weights features in the chest and lungs region more heavily than abdominal regions or the x-ray background. These heatmaps confirm that not only can CNNs correctly classify coccidioidomycosis at a high skill level but they also recognize correct features when making their predictions.

## Discussion

A dataset of 1,174 images is small for deep learning applications. In such a regime, multiple techniques are available to avoid overfitting including: (1) using pre-trained weights; (2) performing various forms of data augmentation; (3) adding various forms of regularization; (4) early stopping as well as using additional randomness in the training procedure, such as dropout (Srivastava, 2014; Baldi, 2014).

The results from all CNN models demonstrate a high level of skill classifying coccidioidomycosis in canine subjects. The high skill level achieved by the models on the unseen test sets demonstrates the models did not overfit and in fact generalized well to new data. Despite the dataset containing canine subjects at different orientations and projections (right lateral, left lateral, and Ventral-dorsal) the models learned to correctly classify coccidioidomycosis in these differing scenarios.

For CNN models to be really useful in a medical capacity they must demonstrate not only high accuracy, but also a good degree of interpretability. The ResNet model probed here demonstrates a high skill level while providing essential interpretability through the use of heatmaps. Coccidioidomycosis is primarily found in the lung fields, the hilar lymph nodes or both. Examining Figure 4, the lung field, located dorsal and caudal of the heart, receives high importance both when predicting the presence of cocci as well as its absence. The lung field cranial of the heart is indicated as important as well.

A critical aspect of the models implemented in this study is the speed at which they compute predictions. The training took place on NVIDA Titan V GPUs with 12GB of memory. Each epoch took roughly two minutes to complete. Training for a typical number of fifty epochs took about sixteen hours for the full 10--fold cross validation. However, at production time, an image can be classified in less than 50 ms, and a heatmap produced in less than 1 second. The low latency offered by these models is further evidence of their potential for being deployed and used alongside medical professionals.

Finally, there is room to further improve the results, for instance by acquiring more data. As the dataset and diversity of samples increases one can expect the performance of CNN models to likewise increase. With more training examples the models will make more accurate predictions and provide more informative localization heatmaps.

## Conclusion

In this paper we presented a novel application of convolutional networks to identify the presence of pulmonary coccidioidomycosis in canine subjects. After assembling a data set of annotated radiographic images, we have trained and compared several deep learning models and demonstrated that successful classification and localization of the disease in radiographic images is feasible. The best model achieves a high skill level and is interpretable via the use of heat maps that can be understood by human experts. While training can be time consuming, at production time the model is fast and can produce useful results in less than a second. As more data is gathered, an automated system driven by machine learning could further improve itself. The same approach could be extended to other organs, diseases, and animal species in veterinary medicine. As is the case for human medicine, the major remaining challenges are in overcoming the numerous barriers to data collection and aggregation, and the actual deployment of deep learning systems in clinical settings where the concerns of multiple stakeholders must be addressed.